\documentclass[conference]{IEEEtran}

\RequirePackage[colorlinks,citecolor=blue,urlcolor=blue,linkcolor=blue]{hyperref}

\usepackage{booktabs}
\usepackage{url}
\usepackage{multirow}
\usepackage{tikz}
\usepackage{pgf-umlsd}
\usepackage{pifont}
\usepackage{tabularx}

\usetikzlibrary{positioning,shapes,shadows,arrows,calc,fit,matrix,chains} 

\usepackage{listings}
\usepackage{graphicx}
\usepackage[tight,footnotesize]{subfigure}
\usepackage{fontenc}

\newcommand{\ohyes}{\ding{51}}
\newcommand{\ohno}{\ding{55}}

\lstdefinestyle{customc}{
	belowcaptionskip=1\baselineskip,
	breaklines=true,
	xleftmargin=\parindent,
	language=HTML,
	showstringspaces=false,
	basicstyle=\footnotesize\ttfamily,
	keywordstyle=\bfseries\color{green!40!black},
	commentstyle=\itshape\color{purple!40!black},
	identifierstyle=\color{blue},
	stringstyle=\color{orange},
}

\lstdefinestyle{customasm}{
	belowcaptionskip=1\baselineskip,
	frame=L,
	xleftmargin=\parindent,
	language=[x86masm]Assembler,
	basicstyle=\footnotesize\ttfamily,
	commentstyle=\itshape\color{purple!40!black},
}

\begin{document}

\title{DOMtegrity: Ensuring Web Page Integrity against Malicious Browser Extensions}

\author{

\IEEEauthorblockN{Ehsan Toreini}
\IEEEauthorblockA{School of Computing Science\\
Newcastle University\\
United Kingdom\\
ehsan.toreini@ncl.ac.uk
}

\and

\IEEEauthorblockN{Maryam Mehrnezhad}
\IEEEauthorblockA{School of Computing Science\\
Newcastle University\\
United Kingdom\\
maryam.mehrnezhad@ncl.ac.uk
}

\and

\IEEEauthorblockN{Siamak F. Shahandashti}
\IEEEauthorblockA{Department of Computer Science\\
University of York\\
United Kingdom\\
siamak.shahandashti@york.ac.uk
}

\and

\IEEEauthorblockN{Feng Hao}
\IEEEauthorblockA{School of Computing Science\\
University of Warwick\\
United Kingdom\\
feng.hao@warwick.ac.uk
}
}

\maketitle

\maketitle

\begin{abstract}

In this paper, we address an unsolved problem in the real world: how to ensure the integrity of the web content in a browser in the presence of malicious browser extensions? The problem of exposing confidential user credentials to malicious extensions has been widely understood, which has prompted major banks to deploy two-factor authentication. However, the importance of the ``integrity'' of the web content has received little attention. We implement two attacks on real-world online banking websites and show that ignoring the ``integrity'' of the web content can fundamentally defeat two-factor solutions. To address this problem, we propose a cryptographic protocol called DOMtegrity to ensure the end-to-end integrity of the DOM structure of a web page from delivering at a web server to the rendering of the page in the user's browser. DOMtegrity is the first solution that protects DOM integrity without modifying the browser architecture or requiring extra hardware. It works by exploiting subtle yet important differences between browser extensions and in-line JavaScript code. We show how DOMtegrity prevents the earlier attacks and a whole range of man-in-the-browser (MITB) attacks. We conduct extensive experiments on more than 14,000 real-world extensions to evaluate the effectiveness of DOMtegrity. 

\end{abstract}


\IEEEpeerreviewmaketitle

\section{Introduction}
\label{introduction}



Browser extensions have become the dominant method to extend browser functionality. All major browsers (Chrome, Firefox, Safari, Opera and Internet Explorer) support extensions, and host dedicated repositories (``stores'') from which extensions can be downloaded and installed directly from the Internet. 
Mozilla reports average rates of more than 1~million Firefox extensions downloaded daily and about 100 new extensions created every day throughout 2017~\cite{addonsStatitics}. 

Extensions are normally distributed and executed in controlled environments. All extensions uploaded to a repository are subject to a \emph{vetting process}, which is a mixture of automated program analysis and manual code review aiming to identify malicious extensions and prevent their spread. Furthermore, extensions are run in a restricted (so-called ``sandboxed'') environment and only have access to a predefined set of browser APIs. 

However, the vetting process is not bullet-proof. A study conducted by Google researchers found nearly 10\% of extensions examined to be malicious~\cite{browserDefence15}. 
By using obfuscation, some malicious extensions can slip through the vetting process. Furthermore, the extension update mechanism provides an additional exploit path for the attacker. In 2014, two popular and previously vetted Chrome extensions, ``Add to Feedly'' and ``Tweet This Page'', were sold to spammers who updated the extensions to inject advertisements and affiliate links into websites opened in the browser. 

\textbf{The Problem}. The key problem with extensions is that, once installed, they possess over-privileged capabilities that may be abused by attackers. For example, an extension is free to modify the \emph{Document Object Model (DOM)} of a web page. This allows a malicious extension to manipulate the display of a web page and deceive users into believing something false. The change of the web page content may be subtle, but when it is combined with social engineering techniques, it can cause significant harm to user security~\cite{adware}. In Section~\ref{motivation}, we will demonstrate two attacks on real-world banking websites (HSBC and Barclays) to show how a malicious extension may stealthily steal money from the user's bank account by making small modifications to the DOM structure of an online banking web page. 

Existing solutions to prevent malicious extensions generally involve changing the browser's internal design~\cite{browserDefence9,sabre,browserDefence3}, strengthening the vetting process of repositories~\cite{browserDefence15,detectionDefence14,detectionDefence4,browserDefence11,browserDefence16}, asking users to install yet another (trusted) extension that detects malicious behaviour of other extensions~\cite{browserDefence1,extensionAttack6} or requiring an external hardware device (e.g., Cronto) that performs out-of-band transaction verification.



\textbf{Our solution}. In this paper, we propose a cryptographic protocol that we call \emph{DOMtegrity} to ensure the integrity of the DOM structure of a web page delivered from a web server to the rendering of the page at the client browser in the presence of malicious extensions. Compared to previous solutions, ours does not require changing the browser's existing internal design; it does not need any external hardware device; it is orthogonal to the strengthening of the vetting process; it can be easily implemented by embedding in-line JavaScript code in the web page rather than requiring the user to install another (trusted) extension. The novelty of our solution lies in exploiting subtle but important differences between extensions and in-line scripts in terms of their rights to access Websockets established between the server and the client. This is combined with leveraging the latest Web Crypto API that is recently added in all major browsers. 



{\bf Contributions. }
The main contributions of this paper are summarized below: 
\begin{itemize}
\item We propose DOMtegrity, a cryptographic protocol to protect end-to-end integrity of a web page's DOM from the point of delivery at a server to the final display in a client's browser. This is the first solution that works with the standard WebExtensions architecture without needing any external hardware. 
\item We present an efficient implementation of DOMtegrity, using JavaScript on the client side and Node.js on the server side, and demonstrate that the proposed solution is effective and only adds a small overhead to the computation load and communication bandwidth.
\item As part of the evaluation, we implement two attacks on real-world online banking systems (HSBC and Barclays) to show how a malicious extension can compromise the security of the user's bank account, and how DOMtegrity can prevent such attacks as well as a whole range of man-in-the-browser~(MITB)~\cite{mitmAttack2} attacks that involve maliciously changing the DOM structure of a web page.
\end{itemize}

\section{Malicious Extension Attacks on Online Banking}
\label{motivation}

Attacks caused by malicious extensions are often known as man-in-the-browser~(MITB) attacks. To demonstrate the importance of understanding the threats imposed by malicious extensions in modern browsers, we show two proof-of-concept attacks on real-world banking websites, HSBC and Barclays, by exploiting the capability of browser extensions to modify the DOM of a web page. The extensions are developed for both Firefox and Chrome based on the standard WebExtensions framework. In the proof-of-concept demonstration of the attacks, the money was transferred between the authors' accounts. All the experiments were approved by Newcastle University's ethics committee. 




\subsection{WebExtensions Capabilities}
\label{WebExtension}

Before describing the attacks, we should first explain WebExtensions\footnote{\texttt{https://developer.chrome.com/extensions/overview}}. The WebExtensions framework is a W3C standard cross-browser architecture~\cite{w3cExtensions} for developing browser extensions using HTML, CSS and JavaScript. It is now supported in all major browsers except Safari. 

An extension developed based on WebExtensions consists of three components: the \emph{background page}, the \emph{UI pages}, and the \emph{content scripts}. The background page is in charge of long-term operations that last beyond the lifetime of a particular browser window and is provided with access to browser APIs. The UI pages put together the extension user interface. Content scripts are JavaScript programs that are run in the context of a web page and are allowed to interact with the page.

Although the background and UI pages do not have access to the DOM of the page, content scripts can modify the DOM. 
Through content scripts, an extension can hide elements of the DOM and insert another element in the same location to effectively replace the original element. For example, a text box can be placed by a malicious extension in place of a password text box to capture a user's password.

\subsection{Attack Model}
In the rest of this paper, the attackers implement their threat scenario through a malicious extension installed in the victim's browser. Thus, the capabilities of a malicious extension are limited to the context of a browser. We assume attackers have not installed any operating system level malicious software on the victim's device to extend their capabilities beyond the browser execution context. 



In the following demonstration, we assume that a malicious extension is already installed on a client's browser. This can be done through disguising malicious extensions as legitimate browser extensions, using Trojans to install such extensions, missing plug-in attacks, or purchasing popular extensions and then adding malicious code during updates~\cite{sessionAttack2,adware}. In both attacks, the web pages that are presented to the victim are from the genuine banking websites via HTTPS.


We assume that the attacker has an account that they wish to move funds to, and the details of this account are either hard-coded into the browser extension or received in real time from a remote Command \& Control (C\&C) center~\cite{perrotta2018botnet}. The attacker's bank account will eventually be exposed by checking the victim's bank transaction records. However, we assume this is not any issue for the attacker since he only needs to prevent the discovery of the fraud for some short timescale in which the funds can be withdrawn from the account.


\subsection{HSBC Attack}
The first attack shows how a malicious extension can easily bypass the two-factor authentication that is adopted by major banks, including HSBC. In this attack, the extension intercepts the victim's authentication credentials (i.e., login details), sends them to a remote attacker and redirects the user to a false maintenance page. Depending on the security policy of the banking web site, this authentication could involve a regular password and an additional one-time password (OTP) as a second factor which is either sent to the user's mobile phone as an SMS or locally generated using a dedicated device (i.e., a Chip Authentication Program (CAP) device) provided by the bank. 

\begin{figure}
	\centering	
	\includegraphics[width=\linewidth]{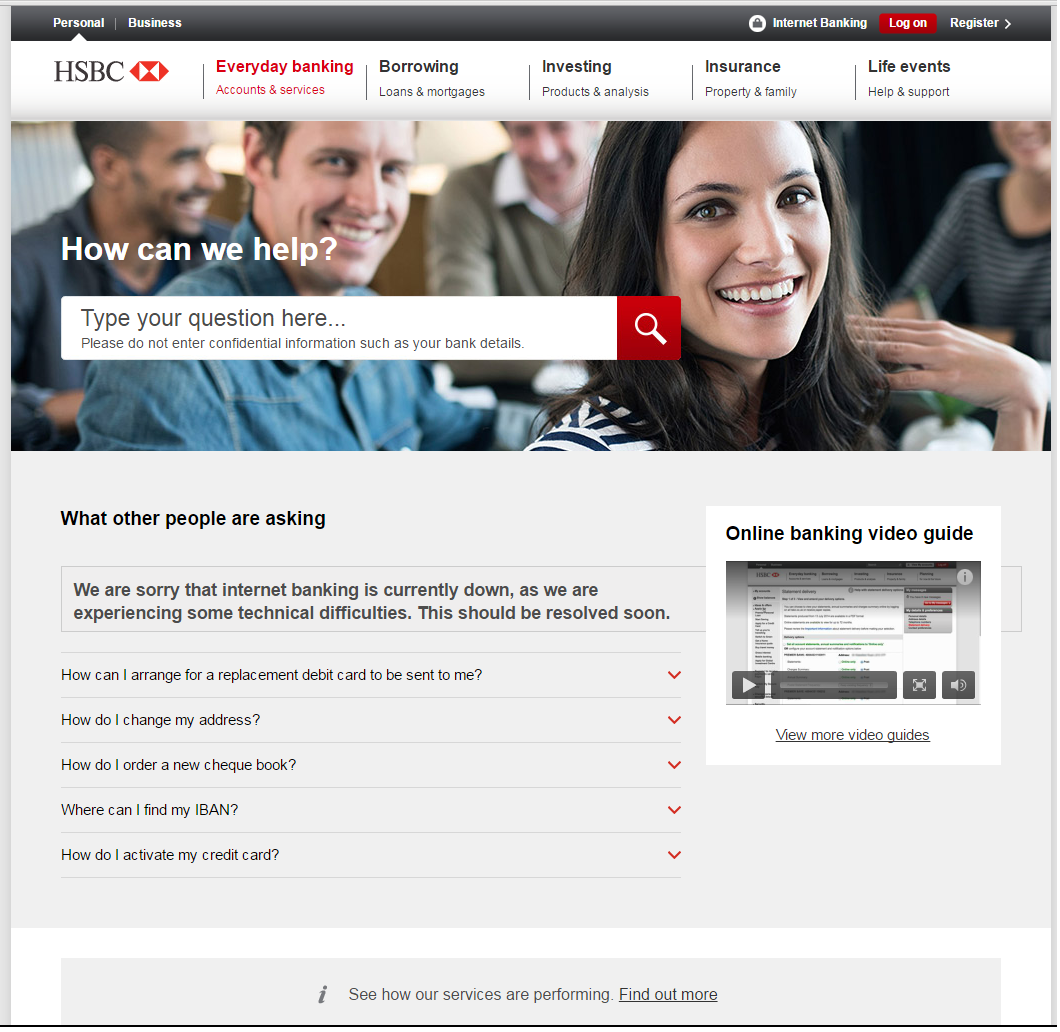}
	\caption{The HSBC customer service page modified by the malicious extension to contain a message indicating website technical difficulties.}
	\label{fig:HSBCAttack}
\end{figure}

We developed a proof-of-concept attack that targets the HSBC online banking web pages. To authenticate their clients, HSBC uses a password-based user authentication augmented with an OTP generated by a dedicated device, the HSBC Physical Secure Key. 
Our attack works as follows: 


\begin{enumerate}

\item When the victim requests the login page, the browser extension content script replaces the username and password text boxes with its own and records the victim's username and password by communicating with the extension background page. 

\item When the victim is prompted for an OTP, the browser extension records what the victim enters in a similar manner. 

\item The victim is then redirected to a genuine customer service page. However, the content of the page is changed on the fly by the extension content script to include a message indicating that the website is temporarily unavailable for maintenance or due to technical difficulties as shown in Figure~\ref{fig:HSBCAttack}. 

\item The stolen login credentials are sent to the attacker who can then log into the victim's online banking account. 

\end{enumerate}

We have implemented the attack by developing extensions for both Firefox and Chrome based on WebExtensions. Our extensions were able to perform the attack successfully without being detected by the bank server. Consequently, we were able to impersonate the victim and log into his or her bank account on a separate machine. 

\subsection{Barclays Attack}

The second attack shows how a malicious extension can defeat transaction-specific user authorization, which is added by many banks such as Barclays as an extra layer of security on top of two-factor authentication. Here, when an already authenticated user requests a transaction, she is required to provide a transaction-specific authorization code which is either sent to the user out of band or generated by a dedicated device upon unique transaction-specific input. This \emph{transaction authentication} is designed to prevent modification of transaction data (e.g., recipient and amount) by man-in-the-browser attackers. 

Barclays uses the strongest form of transaction authentication (the so-called \emph{full transaction authentication}~\cite{extensionAttack3}) in which the unique transaction authorization code (i.e., the transaction-specific OTP) is cryptographically bound to the transaction data. The authorization code is calculated by a dedicated device provided by Barclays called PINsentry. Alternatively, the user can use the functionally equivalent Mobile PINsentry application on her smartphone. PINsentry 
is a battery-powered device consisting of a numeric keypad, a small LCD screen, a card reader and a processor. When a transaction is requested through Internet banking, the user is required to manually enter the transaction details, including the payee account number and the amount, on PINsentry (or Mobile PINsentry) and then enter the PINsentry produced authorization code on the internet banking web page. However, in the following we show how a malicious extension can defeat this security measure by combining social engineering and DOM modifications. The attack works as follows:


\begin{figure}[t]
	\centering	
	\includegraphics[width=\linewidth]{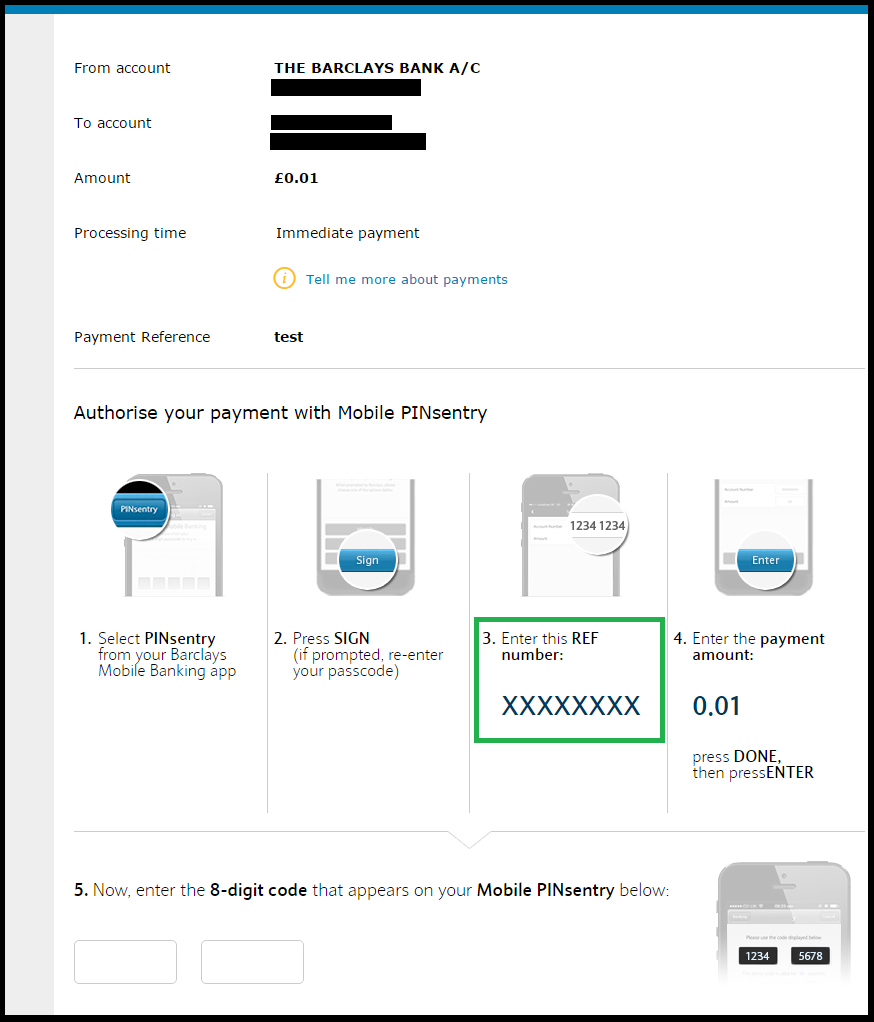}
	\caption{The Barclays instructions page modified by the malicious extension to include the attacker's account number (redacted as XXXXXXXX) as the REF number. The modified area is represented in the green box.}
	\label{fig:barclaysAttack}
\end{figure}
\begin{enumerate}

\item When the victim requests a funds transfer, she is presented a form to provide the details of the funds transfer, including the payee account number and the amount. The malicious extension content script replaces the text box where the victim is supposed to enter the account number of the intended payee with its own text box and records the entered account number by communicating with the extension background page. 

\item Then the user is presented with a dialogue confirming the transaction details and instructing her how to get a transaction authorization code from PINsentry. The instructions include asking the user to ``Enter the payee's account number as your REF:'' followed by the payee's account number. The malicious extension content script replaces this instruction with ``Enter this REF number:'' followed by the attacker's account number, as shown in Step~3 of the instructions in Figure~\ref{fig:barclaysAttack} with real bank details suitably redacted. 

\item A non-expert user, trusting the HTTPS page to be secure and failing to notice the above subtle change, then enters the attacker's bank details in PINsentry and provides a code authorizing the funds transfer to the attacker's account. 

\item The browser extension changes the final confirmation page before it is displayed to the user so that it shows the account details of the original intended payee rather than that of the attacker. 

\end{enumerate}



The key issue that we were able to exploit is that PINsentry prompts the user for two pieces of transaction information: ``REF'' and ``Amount''. The only information about what ``REF'' means is present on the website, which can be modified by the extension. We have responsibly disclosed our attack to Barclays and since then Mobile PINsentry has been updated and the prompt on the app has been fixed to explicitly ask the user for the payee's account number instead of a REF number.






\section{Our Proposed Solution: DOMtegrity}
\label{protocol}

In this section, we propose a solution, called DOMtegrity, to address MITB attacks such as those demonstrated in the previous section. Our solution is designed based on the WebExtensions framework, which is now the standard extension development architecture recommended by W3C and adopted by Google Chrome, Mozilla Firefox, Microsoft Edge and Opera. 




\subsection{WebExtensions Security Model}
\label{sec:WebExtSecModel}
The WebExtensions security model as implemented in modern browsers is based on the model proposed by Reis et al.~\cite{detectionDefence11} who discussed the real-world security issues experienced by Google Chrome and advocated a systematic method to prevent these attacks. Here we discuss parts of this model that are necessary for the description of our protocol. 


\begin{figure}[t]
	\centering	
	\includegraphics[width=1.12\linewidth, scale=0.5, trim={2.5cm 0 0 0},clip]{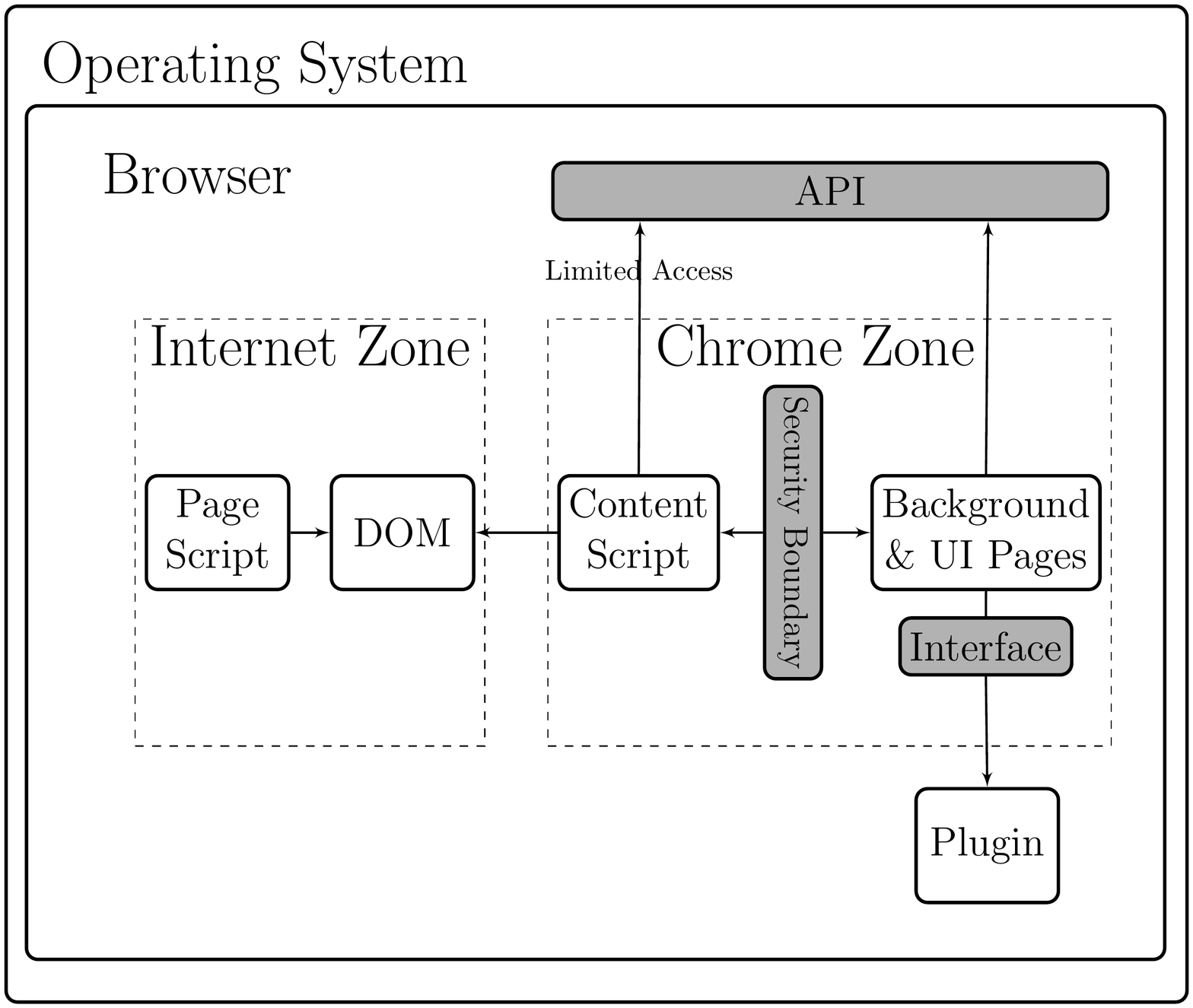}
	\caption{The Internet and Chrome zones of a modern browser and how web pages, extensions, and plug-ins interact~\cite{browserAttackBook}}
	\label{fig:chromeArchitecture}
\end{figure}

{\bf Browser Zones.}
In modern browsers, the execution environment is divided into two zones: an unprivileged \emph{Internet zone} in which web pages are executed, and a privileged \emph{Chrome zone} in which extensions are executed. A schematic representation of these zones is shown in Figure~\ref{fig:chromeArchitecture}. Scripts in the Internet zone (i.e., the so-called \emph{in-line} scripts within the web page) cannot have access to the data in the Chrome zone (i.e., the extension scripts), and vice versa. Therefore, although the web page scripts and the extension content scripts can interact with DOM separately, they cannot interact with each other. This concept is called the \emph{isolated worlds} principle~\cite{isolatedWorld}. The main reason for the isolation is to prevent malicious in-line scripts from exploiting the vulnerabilities that may exist in extension content scripts~\cite{browserAttackBook}. However, as we will explain, the isolation is also useful in defending against malicious extensions when the in-line scripts are from a legitimate source.



\textbf{Permissions.}
Every extension must provide a ``manifest'' in the JSON format which defines the resources and the corresponding \emph{permissions} for each component of the extension. Based on this manifest, users are asked to grant the required permissions at the time the extension is installed, and once installed, the extension's access to browser APIs is limited to these permissions.

\subsection{Design Overview}

DOMtegrity is designed to enable the server to detect any unexpected modification of the DOM by extensions when the web page is rendered in the browser. The underlying idea is that DOMtegrity securely records all the modifications made to the web page DOM until the final rendering of the page and then securely communicates the recorded modifications to the server. The server is then in a position to decide whether or not the client's browser has parsed the page as the server expected. 

DOMtegrity is implemented as a JavaScript program, called \texttt{pid.js}, which is then embedded as an in-line script (within a \texttt{$<$script$>$} tag) in the web page that the server wishes to  protect. This in-line inclusion is necessary since extensions are not able to restrict the execution of in-line web page scripts, whereas they can block loading external script files. For the in-line Javascript to work, we assume that JavaScript execution is not disabled in the browser. 

Since DOMtegrity is to record all modifications to the DOM, it is essential that \texttt{pid.js} is placed at the start of the page source code and before all other HTML tags. Since parsing the web page in browser proceeds in the order that tags are placed in the page source code, placing \texttt{pid.js} at the start of the page ensures that recording changes in the DOM starts immediately as the browser starts parsing the page. 


\begin{table*}[t]
    \caption{\label{tbl:pidandExtension}Capabilities of extension and in-line script (W3C~\cite{w3cExtensions}).}
    \centering
        \begin{tabular}{|l|c|c|}
            \hline
            Capability &  Extension & \texttt{pid.js} \\ \hline
            Access the DOM & \ohyes & \ohyes \\ \hline
            Establish Websockets & \ohyes & \ohyes \\ \hline
            Block Websocket Establishment & 
             \ohno & \ohno \\ \hline
            Block Websocket communications & 
             \ohno & \ohno \\ \hline
            Access an expando created by \texttt{pid.js} & 
             \ohno & \ohyes \\ \hline
            Access/close Websockets established by \texttt{pid.js} &
             \ohno & \ohyes \\ \hline
            Access/close Websockets established by the extension & 
             \ohyes & \ohno \\ 
            \hline
        \end{tabular}
\end{table*}

The isolated worlds principle guarantees that DOMtegrity's recording of modifications in DOM cannot be tampered with by any extension. When executed, \texttt{pid.js} creates an on-the-fly DOM property (also called a DOM expando) named \texttt{document.pid} which implements the DOMtegrity functions within a domain isolated from any extension. 

DOMtegrity uses the recently introduced \emph{Websocket}\footnote{\texttt{https://www.w3.org/TR/Websockets}} technology which provides a full-duplex communication channel over TCP (or SSL/TLS for an encrypted channel) and is now supported by all major browsers. In this paper, we only consider Websocket established over the secure SSL/TLS channels.
The important property here is that although both in-line scripts and extension content scripts can establish Websockets, neither has access to Websockets established by the other. 


 \begin{figure}
	\centering	
	\includegraphics[width=\linewidth, scale=0.4, trim={1cm 1cm 1cm 0cm},clip]{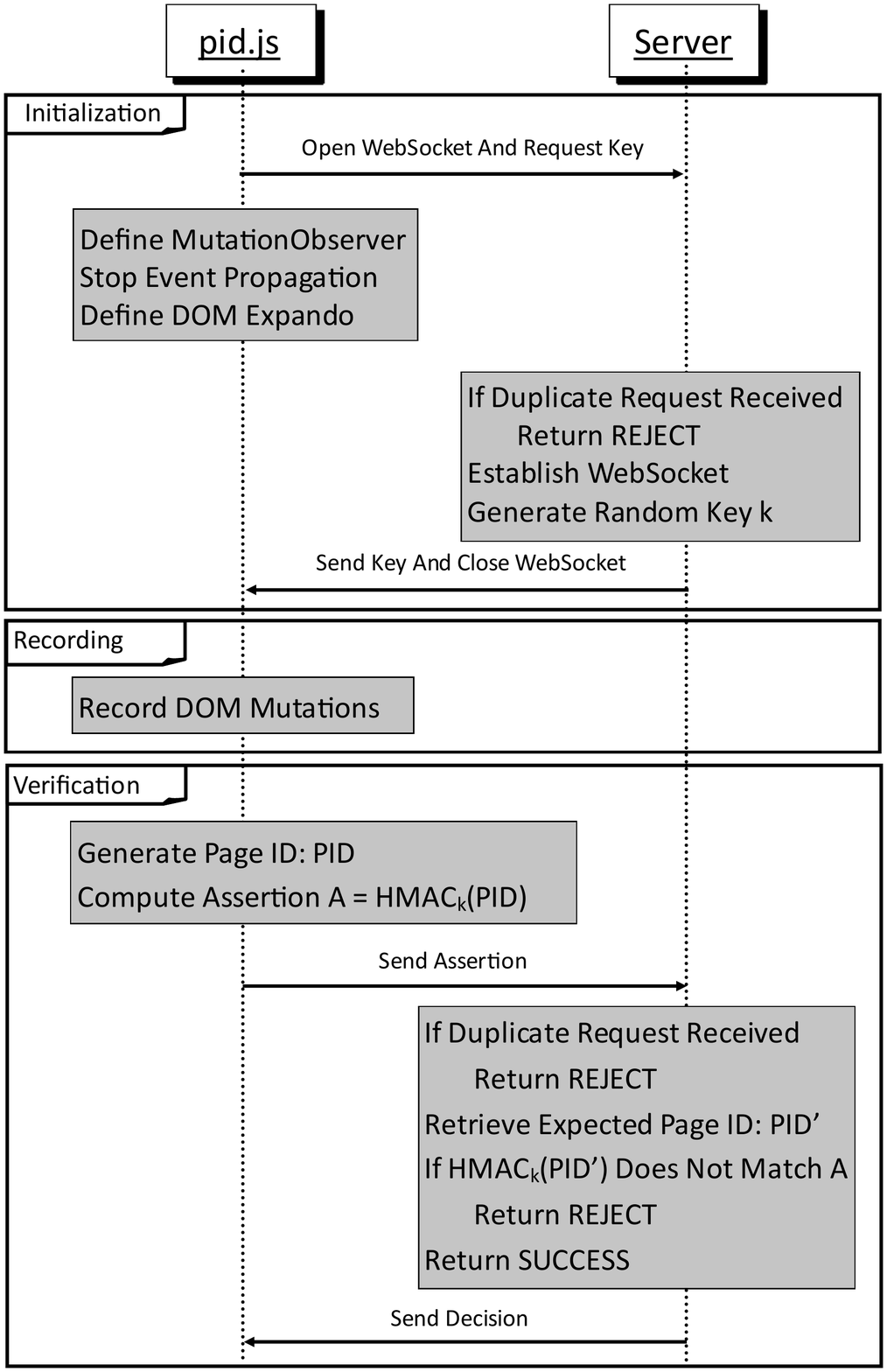}
	\caption{sequence diagram for DOMtegrity}
	\label{sequenceProtocol}
\end{figure}

The extension's inability to access Websocket communication established by DOMtegrity provides assurance on the integrity of the communication between \texttt{pid.js} and the server. The in-line script \texttt{pid.js} establishes a Websocket with the server and this Websocket is used as a secure channel to convey a secret key which is later used to authenticate the DOM modifications that \texttt{document.pid} records. We should emphasize that although an extension has extensive access to HTTP(S) communications, it can only access the Websockets that are established by the same extension. 

Table~\ref{tbl:pidandExtension} summarizes the relevant capabilities of extensions compared with in-line scripts such as \texttt{pid.js} based on the latest W3C specification (dated 23 July 2017)~\cite{w3cExtensions}. Both can access the DOM and establish Websockets, but neither can block Websocket communications. The extension cannot access the expando created by \texttt{pid.js}. Neither \texttt{pid.js} nor the extension can access or close Websockets established by the other. 



\subsection{Detailed Description}
\label{detailedDesign}
DOMtegrity runs in three stages: initialization, recording and verification. The initialization stage sets up the protocol, the recording stage is in charge of storing all DOM modifications, and eventually in the verification stage evidence of DOM integrity is generated on the client side and is sent to the server for verification. These stages are described in detail in the following. A sequence diagram of the protocol is shown in Figure~\ref{sequenceProtocol}. We assume the web page is served over HTTPS. The client is identified by the TLS session ID. 


\subsubsection*{Stage 1: Initialization}
This stage begins as the browser starts parsing the web page. In this stage, the required setup for DOMtegrity is carried out as follows:

{\bf Open Websocket and Request Key.}
First, \texttt{pid.js} sends a request to open a Websocket in order to receive an HMAC key from the server. The server caters for such a request only once within an HTTPS session. To cater for the request, the server establishes a Websocket channel with the client, and through this channel sends a random 256-bit key $k$. The Websocket is subsequently closed and the rest of the communication is continued over HTTPS. Any further requests for a key in the same HTTPS session are refused by the server. If the server receives more than one request for the client, it is an indication that a malicious extension tries to impersonate the client. 

 
{\bf Define Mutation Observer.}
The next step is to assign a \emph{mutation observer}\footnote{\texttt{https://developer.mozilla.org/en/docs/Web/API/{\allowbreak}MutationObserver}} to the document class. Mutation observer is a JavaScript global API that provides developers a way to react to DOM modifications. It records all the changes in the DOM tree, including the alternations in attributes. This covers every possible DOM modification with the exception of the changes in the way events are handled in DOM. We discuss how to deal with this exception below. 


{\bf Stop Event Propagation.}
In this step, \texttt{pid.js} stops assignment of new events to DOM elements by calling the \texttt{stopImmediatePropagation} method\footnote{\texttt{https://developer.mozilla.org/en/docs/Web/API/Event/{\allowbreak}stopImmediatePropagation}} for all elements. Note that (in DOM Level 2 and above) existing assigned events cannot be changed or removed unless the browser is presented with the reference to the registered event, and the isolated worlds principle ensures that extensions do not have access to such references. 



\begin{figure}
	\centering	
	\includegraphics[width=\linewidth, scale=0.5, trim={2.5cm 0 0 0},clip]{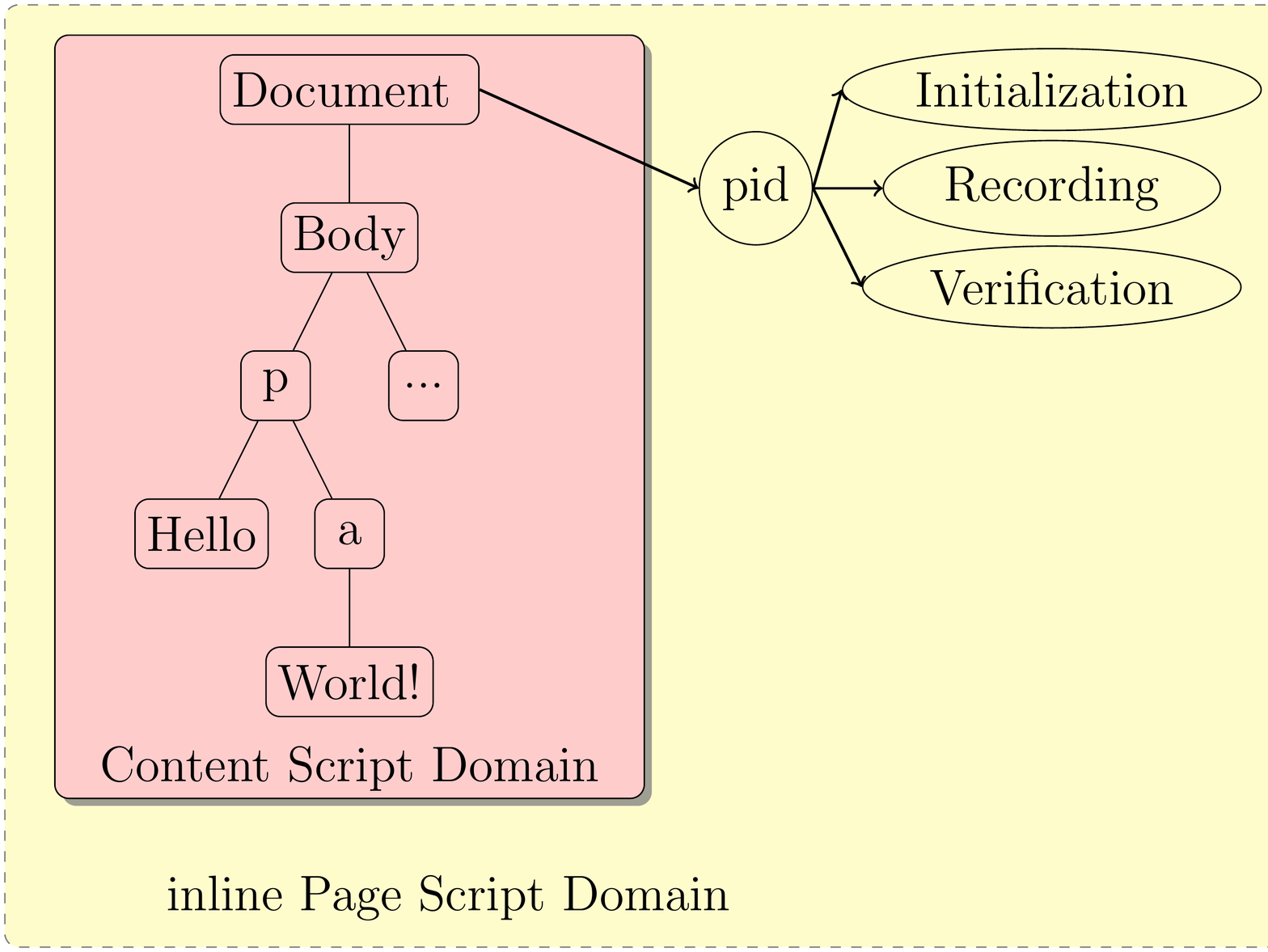}
	\caption{An overview of document.pid and the inability of extensions to modify this region of DOM.}
	\label{fig:DOMExpando}
\end{figure}

{\bf Define DOM Expando.}
Next, the script adds an expando (i.e., an on-the-fly property) to the document node of the DOM, as shown in Figure~\ref{fig:DOMExpando}. This property is called \texttt{document.pid}. As a property it does not change the DOM node structure, and hence is not visible to extension content scripts due to the isolated worlds principle. \texttt{document.pid} is implemented as an object with encapsulated functions. All \texttt{document.pid} functions are private (using so-called ``closures''\footnote{\texttt{https://developer.mozilla.org/en/docs/Web/{\allowbreak}JavaScript/Closures}}) except for one (i.e., \texttt{document.pid.request()}) which we discuss later. 

\subsubsection*{Stage 2: Recording}
After initialization, DOMtegrity enters a persistent passive mode and records all DOM mutations through the mutation observer. The recorded mutations include adding or removing child elements to a node, inserting or changing an attribute in a node, or modifying the data of a DOM node. The recording continues until the user's interaction with the web page finishes and the filled form is to be posted to the server. 

\subsubsection*{Stage 3: Verification}
In this stage, a page identifier (PID) containing the recorded changes in the DOM is generated. The stage starts when the function \texttt{document.pid.request()} is called. This is the only public expando function and should be called when the client ``returns'' the form, e.g., by clicking a ``submit'' button. This stage uses Web Crypto API\footnote{\texttt{www.w3.org/TR/WebCryptoAPI}}, a relatively new JavaScript capability to perform cryptographic operations in browser. 

{\bf Generate Page ID.}
The first step is to generate the PID which consists of two parts: the list of recorded DOM mutations throughout the recording stage, and the source code of the page at the time the verification stage starts. According to the W3C standard, there are seven mutations observable. Each possible DOM mutation is encoded into a unique digit to achieve a short representation of the list. The source code (accessible to JavaScript via the \texttt{document.documentElement.innerHTML} attribute) represents the final state of the DOM elements in the page. Here we consider the protection of integrity for the whole page, but it is possible to define a custom PID to cover only part of the page. 

{\bf Compute Assertion.}
Next, a message authentication code (MAC) on the generated PID is produced in the browser using the secret key $k$. We opted to use HMAC with the SHA-256 hash function as our MAC. This selection is based on two main reasons: first, the 128 bit security of the HMAC-SHA256 is adequate for nearly all practical web applications; second, the HMAC function is supported consistently in all modern browsers.
The computed HMAC tag is sent to the server for verification as an \emph{assertion}. 

{\bf Verify Assertion.}
On the server side, upon receiving the assertion, the server first checks if more than one request for fetching the HMAC key has been received earlier within the HTTPS session, and rejects the assertion if that is the case. Multiple key fetching requests indicate man-in-the-browser impersonation attacks. If only one request has been received, the server retrieves the expected PID, computes the HMAC of the expected PID and compares it with the received assertion. Normally there is no need for the client to send the PID. The server expects no changes in the DOM other than those made by the web page scripts. Hence, the server has a specific expectation of the recorded DOM mutations and the final source code of the page, and therefore a known expected PID. The server accepts the assertion on the integrity of the page if the HMAC verification succeeds. Depending on the decision, the server proceeds to provide or refuse further service to the client. In case of refusal, the server may additionally send an error message through an out-of-band channel, e.g., an SMS message to the user's mobile phone.

In the protocol described above, we assume the legitimate changes of DOM can be pre-determined, hence the server is able to derive an \emph{expected} PID. In this case, the client does not need to send the \emph{actual} modifications to the server. The server can verify the HMAC tag against an \emph{expected} PID to decide acceptance or rejection. However, in some cases the changes of DOM may not be fixed (e.g., they may depend on user interactions). To address this, we only need to slightly modify the protocol by sending PID along with the assertion to the server. This way, the server can first verify the HMAC tag against the received PID, and then examine the changes recorded in the PID according to some rules to determine if they are legitimate or not.

\subsubsection*{Choosing HMAC vs.\ Hash} 
DOMtegrity uses the Websocket to securely transport a key which is later used in the generation of the HMAC tag. The Websocket channel only lasts for the duration of the key transport and is immediately closed by the server once it sends the key. An alternative approach would be to keep the Websocket open for the duration of the protocol and instead of sending an HMAC of the PID, the client can securely send a hash (say SHA-256) of the PID through the Websocket. We chose the HMAC approach to minimize the cost of communication since maintaining a full-duplex Websocket requires exchanges of ping-pong messages to keep the channel alive. 
By using HMAC, DOMtegrity minimizes the duration of a Websocket only for the essential purpose of transporting a short (32 bytes) key. As we will show, the computation of HMAC based on WebCryptoAPI incurs a negligible cost in the client browser. The computed HMAC tag can be sent through an XHR request over HTTPS.


\subsection{How DOMtegrity Prevents Attacks}
\label{impersonation}
In this section, we review a number of design choices in DOMtegrity that are essential to effectively defend against DOM manipulation attacks by malicious extensions. 

\textbf{Influencing the execution of pid.js.} A malicious extension may try to influence the execution of \texttt{pid.js} through the content scripts or the injected scripts. First of all, it cannot stop or change \texttt{pid.js} functions through its content scripts. Due to the isolated worlds principle, and that DOMtegrity procedures are defined as \texttt{document.pid} expando functions, the extension content scripts cannot block or manipulate these procedures.
Furthermore, a malicious extension cannot stop or change \texttt{pid.js} functions through injection of scripts into the page. 
Injected scripts do not have access to the \texttt{pid.js} Websocket due to \emph{closure}. The only interference that injected scripts can cause with DOMtegrity is to call the public function \texttt{document.pid.request()}. However, this will result in the rejection of the integrity assertion since the inject script changes DOM by adding a new \texttt{<script>} tag. 





\textbf{Polluting JavaScript variables.} A malicious extension may inject malicious scripts into the page, trying to pollute the local and global variables used by \texttt{pid.js}. First, because we leverage JavaScript \emph{closure} to make a protected reference to Websocket, an injected malicious script cannot access the local Websocket variable in \texttt{pid.js}. Second, an injected script cannot prevent Websocket establishment by DOMtegrity through redefining global JavaScript APIs~(a process known as ``monkey patching''). The isolated worlds principle prevents extensions from modifying parameters of a page's global environment through content scripts. Hence, the only avenue to modify such global definitions would be injecting scripts into the page. There are two cases here. In the first case, the malicious extension ensures the injected script runs before \texttt{pid.js} (which can be realized by setting \texttt{run\_at} to \texttt{document\_start} in the manifest). However, at \texttt{document\_start} which refers to the time before the DOM is created by the browser engine, there is no DOM for the injected script to insert a \texttt{<script>} object, as a result there is no influence on the parsing of \texttt{pid.js}. In the second case, when the injected script runs after \texttt{pid.js}, DOMtegrity's objects have already been created based on default (clean) variable definitions. In the implementation of \texttt{pid.js}, we leverage the \texttt{Object.freeze()}~\cite{contentIntegrity3} function to freeze the DOMtegrity APIs in the initialization phase, hence making the DOMtegrity object immutable. This prevents an injected malicious script from performing any modifications to the global variables used in \texttt{pid.js} after it is parsed.


\textbf{Eavesdropping the secure channel.} The \texttt{pid.js} Websocket provides a secure communication channel between \texttt{pid.js} and the server. This channel is inaccessible to the malicious extension~\cite{webrequest}. In other words, the extension cannot read or modify data sent through this channel.

\textbf{Impersonation.}
\label{race}
The design of DOMtegrity was based on the W3C standard on ``browser extensions''~\cite{w3cExtensions}. A malicious extension may try to impersonate \textsf{pid.js} by sending a request to establish the Websocket first.
However, according to the W3C specification~\cite{w3cExtensions}, an extension is not allowed to stop \texttt{pid.js} from sending its own Websocket request. The setting of \texttt{document\_start} in the manifest of the extension can enforce the execution of content scripts before parsing the loading page. However, a meaningful attack would need the user to interact with a web page that is loaded in the browser (e.g., to fill in a form or to click a button). The inclusion of \texttt{pid.js} before the web page HTML code ensures that the user interaction can only happen after \texttt{pid.js} sends its own Websocket establishment request. Hence, any attempt for an impersonation attack by the malicious extension is detected at the server side as a result of observing multiple Websocket establishment requests.

\section{Implementation and Evaluation}
\label{evaluation}


In this section, we describe how we implemented a number of proof-of-concept malicious extensions to test our solution in several attack scenarios and provide performance measurements. 

On the client side, DOMtegrity is implemented as a single JavaScript program which is integrated in-line within a \texttt{<script>} tag in the beginning of a web page. On the server side, we implemented the server using Node.js version~4.4.0. 
All cryptographic operations in \texttt{pid.js} are programmed as asynchronous operations using JavaScript \emph{Promise} objects\footnote{\texttt{https://developer.mozilla.org/en/docs/Web/JavaScript/{\allowbreak}Reference/Global\_Objects/Promise}}.

\subsection{Confirming DOMtegrity Effectiveness}


\textbf{Detecting Online Banking Attacks.}
To confirm that our implementation of DOMtegrity can detect the attacks we discussed in Section~\ref{motivation}, we implemented copies of the online banking web pages for both systems on our local server and embedded \texttt{pid.js} in-line. Then, we re-ran the attacks by the malicious extensions we developed on Chrome and Firefox. 
In both cases the server was able to successfully detect the malicious modifications made on the web pages and block further requests from the client. 

\textbf{Detecting Other Possible DOM Modifications.}
To confirm that our implementation of DOMtegrity can detect other possible DOM modifications, we considered a comprehensive list of changes extensions can make to DOM and developed extensions that make such changes through content scripts. These changes include:
%
\begin{enumerate}
\item insert a new DOM element into the tree;
\item remove a targeted DOM element from the tree;
\item hide a targeted DOM element and replace it with its own element (possibly of an identical type) with a different ID;
\item change the style of a targeted DOM element; and 
\item embed another script file which in turn changes an attribute of a targeted DOM element.
\end{enumerate}


We developed five extensions (based on WebExtensions), each making one of the above modifications. All these extensions are tested on a simple login web page, which 
contains username and password text boxes and a ``Sign in'' button, with \texttt{pid.js} embedded in-line. We tested each of our extensions on Chrome and Firefox. As we expected, in all the experiments our server was able to detect the malicious DOM modifications on the client side. 






\subsection{Performance Evaluations}
On the client side, the web page is run in Firefox v50.1 and Chrome v54 on a machine equipped with Intel Core i7 2.8\,GHz with 8\,GB of RAM and Windows~7 Enterprise. 
The server is set up on a machine with windows~8.1 x64 Enterprise Edition equipped with Intel Core i5 2.3\,GHz with 8\,GB of RAM.  

\textbf{File Size.}
The client side JavaScript is 550 lines of code and adds 21.6\,KB in the normal mode and 6.33\,KB in the minified mode to the original web page source code. Our simple login page, the HSBC web page and the Barclays web page are 31.5\,KB, 2.1\,MB and 3.6\,MB, respectively. The overhead of the DOMtegrity client source code is relatively small compared to those of other popular JavaScript frameworks. For example, the popular JQuery framework\footnote{\texttt{https://jquery.com}} adds 84.6\,KB to the web page in the minified mode.
The server side Node.js implementation is 240 lines of code with a size of 4.25\,KB. 

\textbf{Computation load.}
The computation load of the initialization stage is proportional to the number of elements in the web page since the browser needs to stop event registration for every node of the DOM. We measured the time it takes for this step to complete for our own login page 
and for the comparatively richer HSBC and Barclays online banking pages. For each page we ran the experiment 100 times and we report the average here. For our login page, this step took 15.64\,ms on Firefox and 16.53\,ms on Chrome to complete, resulting in an average of 0.71 to 0.75\,ms per DOM element. For the Barclays page, the richest page, this step took 624.76\,ms on Firefox and 839.83\,ms on Chrome to complete, resulting in an average of 0.49 to 0.65\,ms per DOM element. Further details are reported in Table~\ref{tbl:eventRegistration}. 


\begin{table*}[t]
\centering
\caption{Average elapsed times for stopping event propagation in Chrome and Firefox for our experimental web pages}
\label{tbl:eventRegistration}
\begin{tabular}{@{}lrrrrr@{}}
\toprule
 & \#Elements & \multicolumn{2}{c}{Total time (ms)} & \multicolumn{2}{c}{Time/Element (ms)} \\ 
 & & Chrome & Firefox & Chrome & Firefox \\ \midrule
\begin{tabular}[c]{@{}l@{}}Simple \\ login page\end{tabular} 
 & 22 ~~~~ & 16.53 \ & 15.64 \ & {\bf 0.75} \ & {\bf 0.71} \\ \hline
\begin{tabular}[c]{@{}l@{}}Simulated \\ HSBC page\end{tabular} 
 & 987 ~~~~ & 713.68 \ & 485.08 \ &  {\bf 0.72} \ & {\bf 0.49} \\ \hline
\begin{tabular}[c]{@{}l@{}}Simulated \\ Barclays page\end{tabular} 
 & 1283 ~~~~ & 839.83 \ & 624.76 \ & {\bf 0.65} \ & {\bf 0.49} \\ 
\bottomrule
\end{tabular}
\end{table*}

The recording stage only stores an encoding of the DOM change for every DOM modification and incurs a negligible computational overhead. In our experiments, the latency for recording each mutation is 0.005\,ms.

\begin{figure}
	\centering	
	\includegraphics[width=0.85\linewidth, scale = 0.4, trim={5cm 0 4cm 0}]{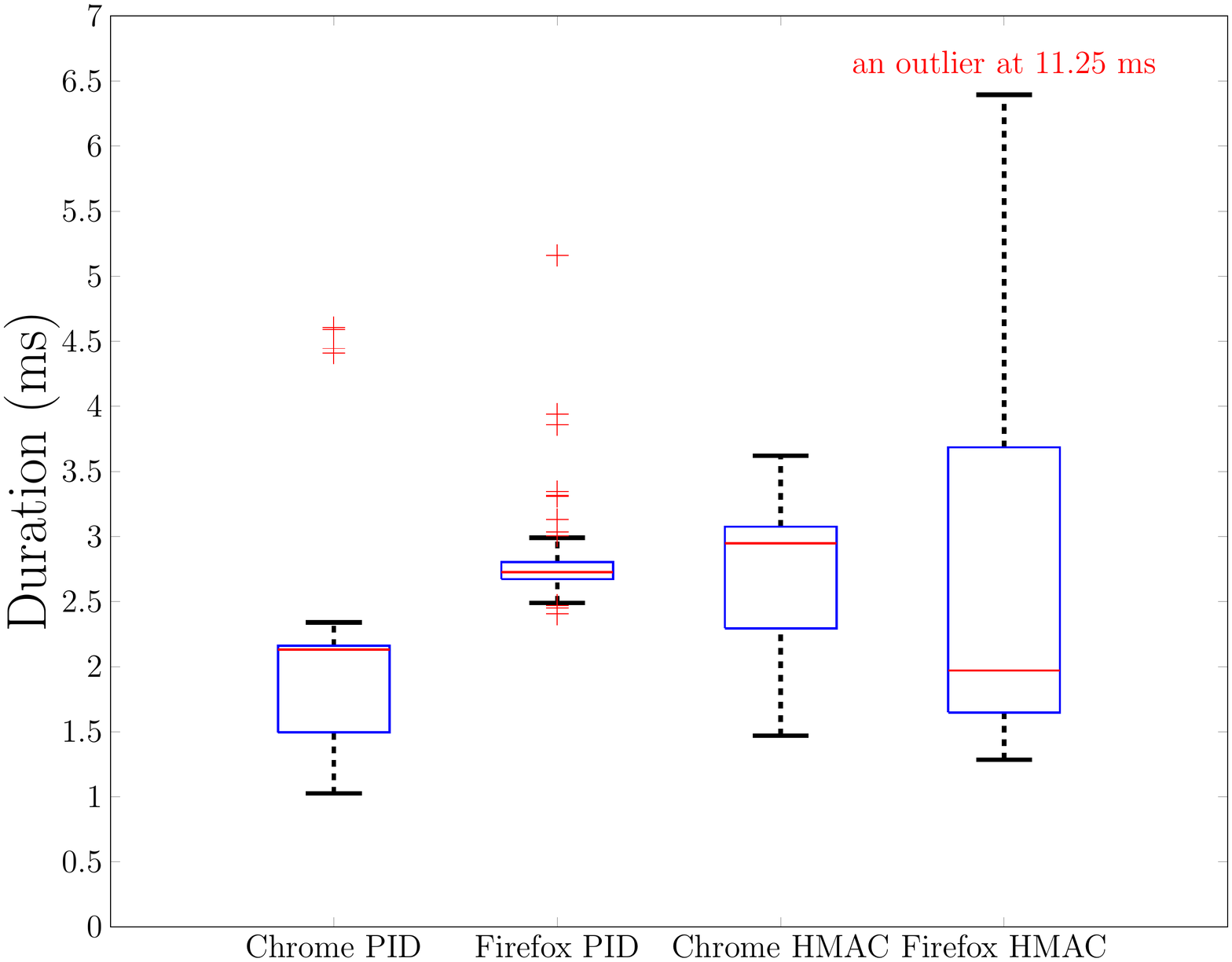}
	\caption{Box plots of elapsed times for PID and HMAC calculations in 100 executions in Chrome and Firefox.}
	\label{fig:performance}
\end{figure}

The verification stage requires the calculation of PID and HMAC tag. In our measurements, the average elapsed time for computation of PID is 1.97\,ms in Chrome and Opera, and 2.79\,ms in Firefox, and the average elapsed time for computing the HMAC tag is 2.63\,ms in Chrome and Opera, and 2.68\,ms in Firefox. The box plots of elapsed times for 100 executions in Firefox and Chrome are illustrated in Figure~\ref{fig:performance}. All values are rounded up to the closest 0.01\,ms. 

Computations on the server side are very efficient. The most time consuming step on the server side is retrieving PID from storage which takes 1.96\,ms on average. It takes 0.17 ms to compute a HMAC tag and another 0.03 ms to compare the tag against the received. The average elapsed time for 100 executions of each step on the server side is shown in Table~\ref{tbl:serverPerformace}. All values are rounded up to the closest 0.01\,ms. 

\begin{table}[t]
\centering
\caption{Average and standard deviation of the elapsed times on the server side for 100 executions of each step of the protocol}
\label{tbl:serverPerformace}
\begin{tabular}{@{}lcc@{}}
\toprule
Step             & Average  time (ms) & STD (ms) \\ \midrule
Key generation   & 0.02 & 0.02 \\ 
PID retrieval    & 1.96 & 1.59 \\ 
HMAC calculation & 0.17 & 0.01 \\ 
Decision         & 0.03 & 0.02 \\ 
\bottomrule 
\end{tabular}
\end{table}

\textbf{Communication Bandwidth.}
DOMtegrity is designed to be efficient in terms of required communication bandwidth. The key and the MAC tag are only 32~bytes each, amounting to a negligible fraction of the usual data transmission between the client and the server. The embedded JavaScript code is relatively compact (21.6\,KB in the normal and 6.33\,KB in the minified mode), as compared to other popular JavaScript frameworks such as JQuery (84.6\,KB in the minified mode). The establishment of the Websocket is also efficient as the underlying technology is designed to be lightweight. By the design of DOMtegrity, the duration of the Websocket channel is kept to the minimum only for the essential purpose of transporting the HMAC key.



\subsection{Compatibility with Real-world Extensions}
DOMtegrity is designed to detect all DOM changes. In the simplest case, when the server is able to anticipate all DOM changes, \texttt{pid.js} only needs to send back a short HMAC tag, which the server can verify against the anticipated changes. However, this may not work with existing real-world extensions that work by modifying the DOM. Examples of such extensions include Grammarly (a popular grammar and spell checker) and LastPass (a popular password manager). In this section, we investigate the compatibility of DOMtegrity with \emph{real-world} extensions. 

\textbf{Real-world extension set.} 
For this experiment, we have downloaded a large set of extensions from the Chrome Web Store and the official Mozilla Add-on repositories. 
Overall, we investigated more than 14,000 WebExtensions-based extensions in the two repositories, as follows: 


\begin{itemize}
\item all extensions from Chrome's Starter Kit list,
\item all extensions from Chrome's Editor Picks list,
\item all extensions returned with the search keyword ``block'',
\item all extensions returned with the search keyword ``blocker'',
\item all extensions with more than 100 active users in each Chrome Web Store extension category, and
\item all WebExtension-based add-ons in Mozilla's top 1,000 most popular extensions (57 extensions). 
\end{itemize}

We installed each extension in a mint instance of the browser, then we requested a DOMtegrity-protected web page, i.e., a page in which the \texttt{pid.js} script was embedded. When the page was completely loaded in the browser, we recorded the generated PID in the presence of the extension on the client side, plus the assertion verification result on the server side. 


\textbf{Results.} We compared the generated PID on the client side with the expected PID on the server side for each rejected extension in order to investigate the type of modification they applied. The W3C specification on DOM categorizes page mutations into three groups: \emph{attributes}, \emph{characterData} and \emph{childList}~\cite{mutationObserverDOM}. The attributes category includes mutations involving modifications of attributes of existing nodes. CharacterData refers to mutations that change any data between the opening and closing tags of a text node. Finally, ChildList includes mutations that involve insertion or removal of nodes in the DOM tree. We investigated the generated PID on the client side and classified the rejected extensions into the above categories. A rejection by the server may be caused by a mixture of the mutation types. In that case, the PID records every type of the mutations.

Overall, 15\% of the extensions caused rejection of the assertion. In other words, 15\% the extensions we collected from the web store modified the DOM. 
Among the 15\% rejections, 86\% of them involved attribute mutations, 2\% characterData mutations, and 98\% childList mutations. 
If we simply record every mutation caused by the extension in the PID, the percentage of occurrence for each of mutations types for attribute, characterData and childList mutations was 43.9\%, 0.2\% and 55.9\% respectively. It would be interesting to investigate if the DOM modification made by the 15\% extensions contain any malicious intent (which we plan to do in future research). Normally, Google quickly removes extensions from the Chrome web store as soon as they are reported to contain malicious code. In the latest example, a malicious Chrome extension, called Droidclub, was removed by Google in February 2018 (along with 88 other malicious extensions)~\cite{Droidclub}, after it had affected half a million users. Droidclub works by injecting a malicious script, hence it falls within the category of childList mutations. Note that our attacks on online banking systems lie in the CharacterData category since the extension changed the fields within a text node.

\textbf{Possible mitigations.} One possible mitigation strategy to accommodate existing extensions 
is to consider a more flexible policy on DOM modifications. 
This will require \texttt{pid.js} to send the PID to the server along with the assertion. The PID consists of the recorded mutations and the final source code. The server can then check the PID against a set of policies to decide if the mutations are acceptable. Thus, further compatibility can be gained by the client sending more data (i.e., the PID) and the server performing slightly more complex verification. 

The above solution also works with dynamic web pages where the DOM modifications depend on how the user interacts with the web page. Such interactions cannot always be anticipated by the server, but can still be checked by the server against rules later once a record of the DOM modifications is obtained.  





\section{Further Discussion}
\label{discussion}

\begin{figure*}
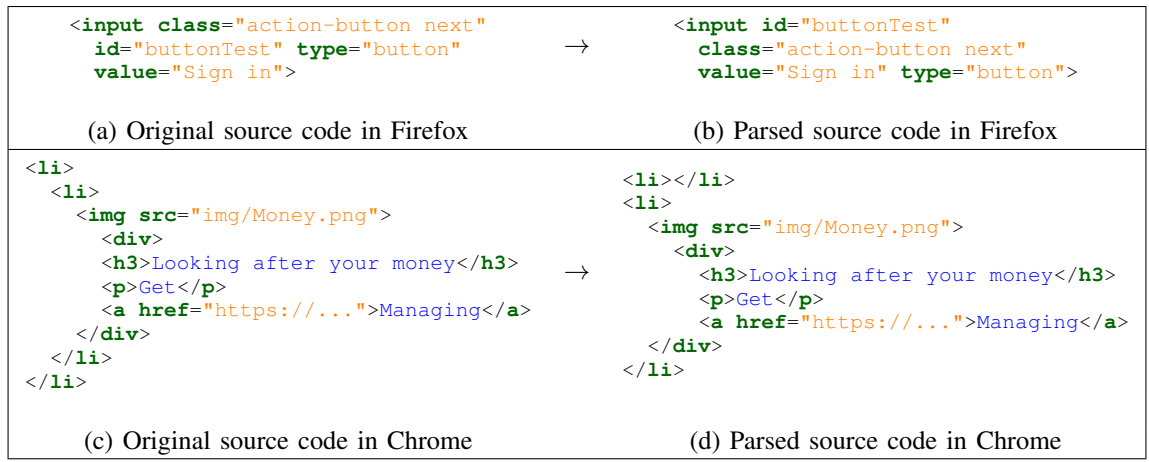

	\centering
      \begin{tabular}{|ccc|}
          \hline
\begin{lstlisting}[language=HTML, style=customc]
<input class="action-button next" 
  id="buttonTest" type="button" 
  value="Sign in">
\end{lstlisting}&
          $\rightarrow$ &
\begin{lstlisting}[language=HTML, style=customc]
<input id="buttonTest" 
  class="action-button next" 
  value="Sign in" type="button">
\end{lstlisting} \\ 
          & & \\
          (a) Original source code in Firefox & & (b) Parsed source code in Firefox \\ \hline

\begin{lstlisting}[language=HTML, style=customc]
<li>
  <li>
    <img src="img/Money.png">
      <div>
      <h3>Looking after your money</h3>
      <p>Get</p>
      <a href="https://...">Managing</a>
    </div>
  </li>
</li>
\end{lstlisting}&
          $\rightarrow$ &
\begin{lstlisting}[language=HTML, style=customc]
<li></li>
<li>
  <img src="img/Money.png">
    <div>
      <h3>Looking after your money</h3>
      <p>Get</p>
      <a href="https://...">Managing</a>
  </div>
</li>
\end{lstlisting} \\
          & & \\

          (c) Original source code in Chrome & & (d) Parsed source code in Chrome \\ 

          \hline
      \end{tabular}

      \caption{Examples of source code modifications during parsing in browsers. Note that modifications observed in Firefox (a) do not apply to Chrome, and modifications in Chrome (b) do not apply to Firefox.}
      \label{fig:parsingChanges}
\end{figure*}

\textbf{Browser Parsing Inconsistencies.}
During the testing of our protocol, we observed two unexpected and undocumented DOM changes made by the browsers in Figure~\ref{fig:parsingChanges}. These changes are caught by DOMtegrity because they modify the source code of the web page. These modifications do not alter the content of the page, but they change the DOM structure. Such changes are harmless from a security perspective, but they are unnecessary and inconsistent between browsers. We reported these minor issues to W3C and Google, and were advised that these appeared to be implementation bugs in the browsers and should be fixed in future releases. This finding shows that although DOMtegrity is designed to detect malicious tempering of DOM, it is also useful to uncover browser implementation bugs. 


\textbf{Dynamic Web Pages.}
A dynamic web page is one with variable content depending on the user or her actions. This is done by either server-side or client-side scripting, or a mixture of both. 

If only server-side scripting is used, a web page is constructed on the server side at the time of request and transmitted to the client. No further changes to the DOM are expected in this case. Hence, such pages can be protected using DOMtegrity as it is designed.

If client-side scripting is used, the dynamic web page DOM is modified in-browser based on the user's interactions with the page. In this case, there would be no way for the server to predict user's interactions with the page and hence it would be necessary for \texttt{pid.js} to send the PID along with the HMAC tag to the server so that a decision on the integrity of the page can be made based on the server's policies. 



\textbf{Private Mode.}
Extension availability policies in private mode are different across browsers. Firefox permits extensions to function in private mode. In contrast, Chrome disables the extensions by default in its private mode (incognito). In each case, DOMtegrity functions as normal, regardless if the extensions are enabled in the client browser.

\textbf{Enabling JavaScript.}
Our solution requires that JavaScript is enabled in the user browser. Obviously it will not work if JavaScript is disabled (e.g., manually by the user, or by setting the CSP response header). In fact, when JavaScript is disabled in the browser, any web page with embedded JavaScript code will stop working. In practice, there are standard techniques to detect if JavaScript is enabled in the browser, and deliver JavaScript-rich content only when it is enabled. The same techniques would apply to DOMtegrity.

\textbf{Confidentiality of data.}
Doomtegrity is designed to protect the integrity of the DOM structure as it is rendered in the browser, but it cannot guarantee the confidentiality of data. A malicious extension is able to read the content of DOM elements as well as http(s) traffic data (and may send the stolen credentials to an external party). This is a privileged capability explicitly permitted by the browser, which treats a browser extension as a ``trusted'' part of the browser~\cite{w3cExtensions}. While our work presents a way to address the integrity problem caused by malicious extensions, we leave it to the future work to address the confidentiality problem, which may require fundamental changes in the browser architectural design.

\section{Related Work}	
\label{related}


This section reviews related work on countering the threats imposed by malicious browser extensions. Existing countermeasures can be categorized into four types: modifying browsers, strengthening the vetting process, requiring another trusted extension and using external hardware. 


%
\textbf{Modifying Browsers}. 
Proposals in this category require their system to be integrated natively within the browser. 
Ter Louw et al.\ design systems for protecting code integrity and user data~\cite{browserDefence9}. The latter is a mechanism that augments the browser to support policy-based runtime monitoring of extension behaviour. The goal is to protect sensitive user data from being accessed or modified by the extension. 
Dhawan et al.\ proposed ``Sabre'', an in-browser information flow monitor to detect malicious activities of JavaScript based extensions during runtime~\cite{sabre}. Sabre associates an appropriate label to all in-memory JavaScript objects based on whether they carry sensitive information. Then, it monitors the objects carrying sensitive information for any insecure access. 
Wang et al.\ proposed an extension access control framework~\cite{browserDefence3}, which dynamically analyses the behaviour of extensions at runtime and controls policies to restrict their access to resources. 
All the proposals in this category require modification of browser code base. Unfortunately none of these proposals have been adopted by mainstream browsers so far. In fact, some of these proposals are based on the XPCOM model for creating extensions in Firefox which is due to be deprecated in favour of WebExtensions.

\textbf{Strengthening the Vetting Process}. 
Proposals in this category involve various techniques to improve detection rates of malicious extensions during the vetting process.
Jagpal et al.\ shared their three years of experience in fighting with malicious browser extensions in Chrome Web Store~\cite{browserDefence15}. They developed a detection system called WebEval to vet the extensions in the market. WebEval combines both static and dynamic analysis of the source code, as well as taking into consideration of the reputation of the extension's developer, and involving human experts in manual reviews whenever necessary. Their method was able to identify real-world malicious extensions with a success rate of 96.5\%. 
%

Besides methods adopted by the industry, academic researchers also propose various techniques to strengthen the vetting process. Kashyap et al.\ proposed a framework to automate the vetting process in official extension repositories~\cite{detectionDefence14}. They proposed a notion of add-on security signature which provides detailed information on its data flow and API usages.
Kapravelos et al.\ presented Hulk as a dynamic analysis system to detect malicious extensions~\cite{detectionDefence4}. They monitored the execution and network activities of extensions to detect their malicious intentions. The had an extensive collection of real-world extensions from Chrome Web Store, and one of their findings was discovering a malicious extension that affected 5.5 million users.
Guha et al.\ proposed an IBEX framework for authoring, analysing, verifying, and deploying secure browser extensions~\cite{browserDefence11}. They suggested a high level programming language to develop extensions. They also proposed Datalog to specify fine-grained access control to restrict the extension's access to security-specific web content. 
Bandhakavi et al.\ presented the VEX framework for highlighting potential security vulnerabilities in browser extension~\cite{browserDefence16}. They applied static information-flow analysis to catch malicious JavaScript code in the extension implementation. 

\textbf{Requiring another Trusted Extension}. 
Proposals in this category require users to trust one particular extension and install it consciously.
Marouf et al.\ proposed a run-time framework called REM that monitors the access made by extensions and provides customized permission~\cite{browserDefence1}. They developed an extension for monitoring other extension based on REM. They monitored API calls from an extension to the browser and enforced their policies on the extension. They notified users about the latest activities of other extensions and allowed them to block future such activities.
Liu et al.\ demonstrated the same threat in Chrome~\cite{extensionAttack6}. They also implemented an extension to enforce more fine-grained privileges to extensions in Chrome. They proposed HTML elements to use another attribute called ``sensitivity'' to differentiate DOM elements and enforce the policy that they call micro-privilege management.

\textbf{Using External Hardware}.
Cronto\footnote{\texttt{https://www.vasco.com/products/two-factor-{\allowbreak}authenticators/{\allowbreak}crontosign.html}} is a commercial hardware-based solution to address MITB attacks specifically for online banking. It was initially developed by a spin-off company from the University of Cambridge in 2005 and was later acquired by VASCO Data Security International for \pounds 17m in 2013. The product has been widely deployed by major banks in Chile, Switzerland and Germany to secure online banking. The Cronto solution works by using a special client device, which shares a secret key with the sever. When the user performs transactions during online banking, the server sends a 2-D barcode to display on the client's web page, which encodes the encrypted transaction details such as the amount, timestamp and account number. The 2-D barcode is then read and verified by the Cronto device that has the decryption key. Upon successful verification, Cronto generates a one-time password (OTP), which the user can enter in the browser to authenticate the transaction. Here, the Cronto device can be either custom-built hardware with an embedded camera or a smart phone.

DOMtegrity is similar to Cronto in preventing malicious modifications on the client side against MITB attacks. However, ours is a JavaScript-based \emph{software} solution and does not require an external hardware token. We note that although the main design aim of Cronto is to ensure the integrity of transactions, it has a secondary function as a second-factor for authentication since the device has a shared secret key with the server. DOMtegrity does not have this function, but it can be used in combination with any existing two-factor authentication scheme, e.g., the Chip Authentication Program (CAP)  currently used by HSBC and Barclays.

\textbf{Other Related Work.}
Reis et al.\ proposed the idea of ensuring web content integrity by JavaScript~\cite{contentIntegrity1}. Their method was inspired by the Linux integrity check and AEGIS~\cite{contentIntegrity2}. The authors developed a client-side JavaScript framework named TripWire, which detects unexpected modifications done by ISPs and other intermediate nodes over HTTP communication. Once the page rendering is complete, the code requested the page's source code from the server through AJAX requests, then the internal source code is compared with the server's one at the client side. Tripwire did not consider browser extensions in their attack model because it considers them as ``trusted''. They discussed that their method was comparable to HTTPS with better performance. Patil~\cite{DOMmutation} proposed another method to isolate DOM from content script. They used \emph{shadow DOM} to present an encrypted view of the page data to the content script. They developed a proof-of-concept prototype in their research.

\section{Conclusion}
\label{conclusion}
In this paper, we present DOMtegrity, a 
JavaScript based solution to provide end-to-end protection of integrity for web content from the point of delivery at a sever to the final rendering in a client's browser. Our solution works with the standard WebExtensions framework and does not require modifying existing architectures of web browsers, nor using any external hardware device. As part of the evaluation, we implement two attacks on real-world online banking websites: HBSC and Barclays, to demonstrate how malicious extensions can compromise the online banking security, and how DOMtegrity can effectively prevent such attacks as well as other man-in-the-browser attacks caused by malicious extensions. We run an extensive study of the top 14,000 extensions to investigate the prevalence and types of DOM changes. Our study confirms that DOMtegrity is compatible with the vast majority of widely-used extensions, and can be made compatible with other extensions after small modifications. We present detailed timing measurements to show that DOMtegrity is efficient and adds only a relatively small overhead to the performance on both the client and the server sides.

\section{Compliance with Ethical Standards}
In this research, we have designed cyber attacks on personal banking websites. These experiments have been performed against the author's bank accounts and the money was only transferred between these accounts. All the experiments were approved by Newcastle University's ethics committee.

\section*{acknowledgements}

We thank the Mozilla browser extension development team for their feedback on our protocol design, state of the art technologies in browser extensions and compatibility of DOMtegrity with their vision of the extension architecture. We thank Jake Arhibald from Google Chrome, Tobie Langle from W3C and Steven Murdoch for many useful comments, and Dylan Clarke for helping implement the extension attacks. An open source implementation of DOMtegrity is  available at \url{https://github.com/toreini/DOMtegrity}. 
This work was funded by the ERC starting grant, No.~306994.


\bibliographystyle{IEEEtran}
\bibliography{extensions}


%
				
\end{document}